\documentclass[showpacs,onecolumn,amssymb, amsmath,nobibnotes, aps,prl,secnumarabic]{revtex4}
\usepackage{bm}
\usepackage{natbib}
\usepackage{graphicx}
\usepackage{amssymb,amsmath}

\begin{document}
\textfloatsep 10pt

\title{Self-organization in turbulence as a route to order in plasma and fluids}
\author{M.G. Shats}
\email{Michael.Shats@anu.edu.au}
\author{H. Xia}
\author{H. Punzmann}
\affiliation{Research School of Physical Sciences and Engineering,\\
Australian National University, Canberra ACT 0200, Australia}
\date{\today}

\begin{abstract}
Transitions from turbulence to order are studied experimentally in thin fluid layers and magnetically confined toroidal plasma. It is shown that turbulence self-organizes through the mechanism of spectral condensation. The spectral redistribution of the turbulent energy leads to the reduction in the turbulence level, generation of coherent flow, reduction in the particle diffusion and increase in the system's energy. The higher order state is sustained via the nonlocal spectral coupling of the linearly unstable spectral range to the large-scale mean flow. The similarity of self-organization in two-dimensional fluids and low-to-high confinement transitions in plasma suggests the universality of the mechanism.
\end{abstract}


\maketitle

Turbulence determines properties of many natural and laboratory systems. Since its effects are often undesirable, the search for routes from turbulence to order is of great practical importance. Turbulence in two-dimensional (2D) fluids self-organizes through the mechanism of the inverse energy cascade. In the absence of dissipation this leads to the condensation of the spectral energy at the largest scale allowed by the system \cite{Krai67}. The condensate formation in 2D fluids has been observed in experiments and numerical simulations \cite{Som86,Paret98,Hoss83,Smith93,Tab02}. It is formed when the dissipation scale, which represents a balance between the spectral transfer and the linear friction, exceeds the system size \cite{Tab02}. In the plasma drift-wave turbulence, the cascade also produces spectral regions similar to inertial ranges in 2D turbulence and leads to spectral condensation \cite{Has79,Fyfe79,Hase87,Hort94,Xia03}. Though main ingredients of the condensation process are known, there is no clear understanding how it is realized in magnetically confined plasma. Here we perform the first comparative analysis of experiments in thin fluid layers and plasma. The processes in 2D fluid turbulence are compared with those in plasma turbulence studied in toroidal magnetically confined plasma of the H-1 heliac \cite{Ham90}. We show for the first time that spectral condensation in plasma turbulence coincides with a state transition, which is similar to low-to-high confinement transitions in tokamaks and stellarators \cite{Wag82,Con00,Fuji03}. The condensation is observed as a metamorphosis of turbulence into a coherent flow. The flow absorbs turbulent energy of the system and enforces highly ordered motion of particles and reduces diffusion. The global coherent flow, seen as a vortex whose size is limited by the system boundaries, is sustained in the higher-order state by the energy supply directly from the linearly unstable spectral region. This mechanism of self-organization is crucial for understanding physics of improved confinement in magnetized plasma. It is also relevant in other 2D turbulent systems and may trigger useful analogies in physics of the Bose-Einstein condensation of quantum gases.

\begin{figure}
\includegraphics[width=16 cm]{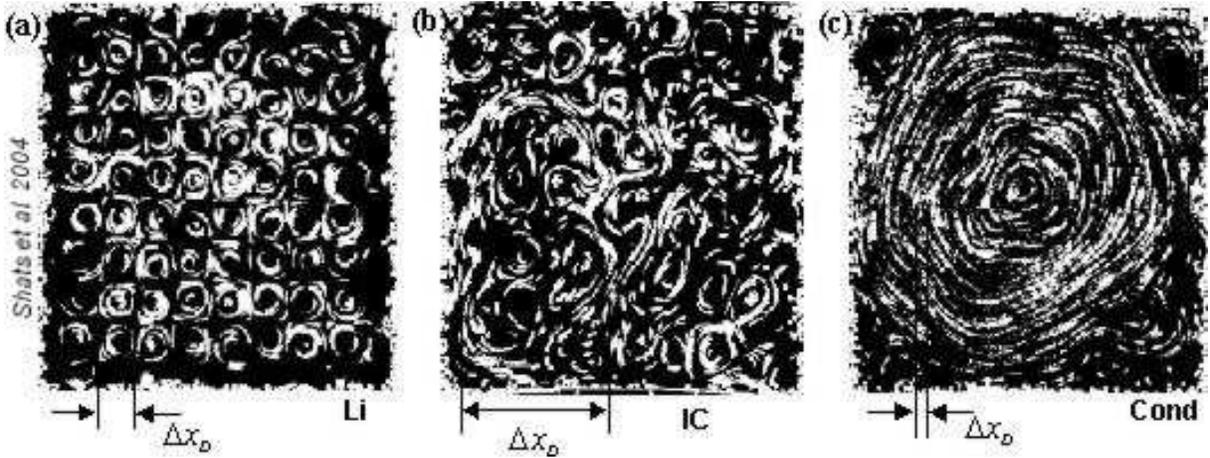}
\caption{\label{fig_1} Evolution of turbulence in 2D fluid. Trajectories
of the tracer particles averaged over 12 frames of recorded video
are shown. (a) The initial (linear) stage, \textit{t}=3~s. (b) The
inverse cascade stage, \textit{t}=25~s. (c) The condensate stage,
\textit{t}=60~s.}
\end{figure}

\begin{figure}
\includegraphics[height=11 cm]{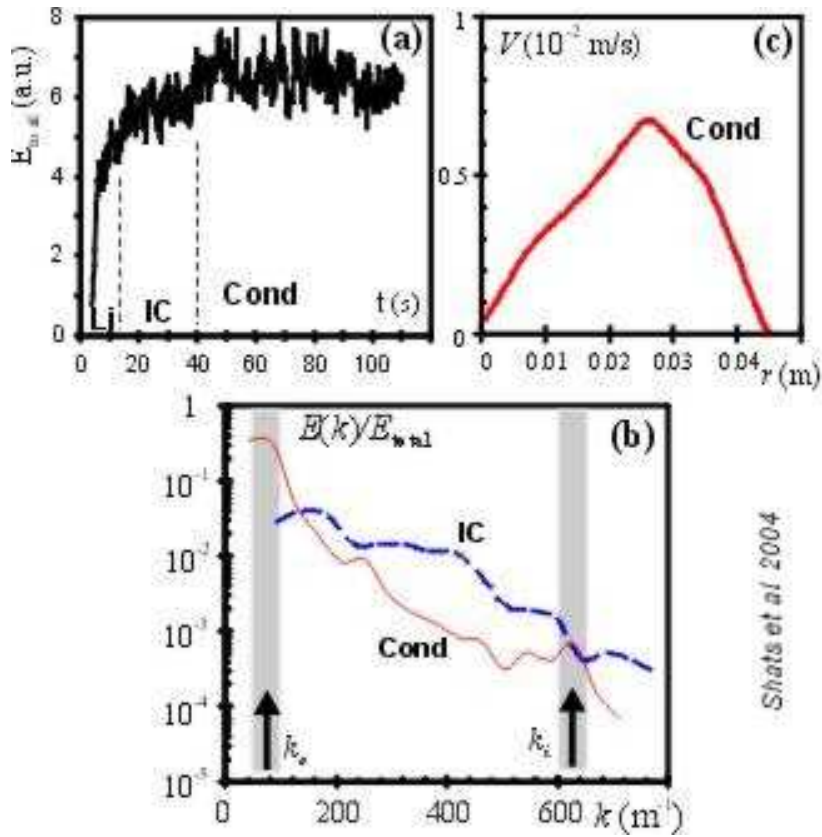}
\caption{\label{fig_2} Analysis of the velocity field during
spectral condensation in 2D fluid. (a) Temporal evolution of the
total kinetic energy: linear (L), inverse cascade (IC) and
condensate (Cond) stages. (b) Fluid turbulence spectra in the
inverse cascade (dotted line) and the condensate (solid line)
regimes. Spectral energy is normalized by the total kinetic energy
of the system. (c) Linear velocity in the condensate vortex as a
function of the distance from the vortex center.}
\end{figure}

First we present results of 2D fluid experiments. To generate the spectral condensate in a fluid we used the experimental procedure reported by Paret and Tabeling \cite{Paret98,Paret97} but with constant instead of random forcing. A turbulent flow is generated in a thin layer of electrolyte (NaCl solution). An electric current driven through the $0.1 \times 0.1$m cell interacts with a $10 \times 10$ matrix of permanent magnetic dipoles placed below the bottom of the cell. The $J \times B$ force generates 100 vortices each of about 10 mm in diameter (1/10 of the box size), thus determining the injection wave number of about $k_i=630$~m$^{-1}$. Magnets are arranged such that any two adjacent vortices counter-rotate.  Since the spectral condensation is very sensitive to the frictional damping, two layers of electrolyte of different concentration (heavier solution at the bottom, total thickness of 6 mm) are used to reduce damping, as proposed in \cite{Paret97}. The flow is visualized using small latex particles floating on the free surface of the electrolyte. Trajectories of the tracer particles are shown in Fig.~\ref{fig_1}. After the force is applied, three stages of the flow evolution are observed. At first, only vortices at the injection scale are seen (Fig.~\ref{fig_1}(a)). As the inverse energy cascade develops, the aggregation process drives larger structures (Fig.~\ref{fig_1}(b)). After about 50~s, a global rotation dominates the flow (Fig.~\ref{fig_1}(c)) and persists in a steady state. The total kinetic energy of the system increases from the linear stage to the inverse cascade stage until it reaches maximum and stabilizes in the condensate regime (Fig.~\ref{fig_2}(a)). Energy spectra of the velocity field are substantially modified during the system's evolution from the inverse cascade regime to the condensate stage (Fig.~\ref{fig_2}(b)). As the system evolves, the turbulent energy is reduced over most of the $k$-range, except for the injection wave number $k_i \approx 630$~m$^{-1}$ and the smallest wave number $k_c \approx 70$~m$^{-1}$ which is determined by the size of the cell. It is this lowest mode at $k_c$ that is seen as the global rotation, or the condensate. A radial profile of the linear velocity of this largest vortex is shown in Fig.~\ref{fig_2}(c).
The establishment of the largest vortex greatly reduces the tracer particle diffusion in the system. As shown in Fig.~\ref{fig_1}, the diffusion scale, $\Delta x_D$, increases from the linear instability stage to the inverse cascade regime. After the global flow is established, the diffusion scale drops dramatically, inhibiting the particle transport from the centre of the largest vortex to its edge and thus reducing diffusion.

To analyse the condensation in plasma, the spectral energy transfer during the turbulence evolution needs to be studied. Such analysis is relatively easy to perform if a single-field description of plasma turbulence is valid. The problem of the mode-coupling in plasma can be reduced to a singe-field model if the electron response to the potential fluctuations is adiabatic. In this case plasma turbulence is described by the Charney-Hasegawa-Mima (CHM) model \cite{Hase78,Hort94}, which also describes atmospheric turbulence. In the reported experiments in \mbox{H-1} at large gyroradii (due to high argon ion mass and low magnetic field), the polarization drift nonlinearity dominates so that the CHM model is applicable to our experiments \cite{Xia03}. This gives the opportunity to: (a) study spectral transfer in a single-field model \cite{Ritz86}, and (b) justify the comparison of the spectra evolution in 2D fluid experiments with plasma results.

\begin{figure}
\includegraphics[height=11 cm]{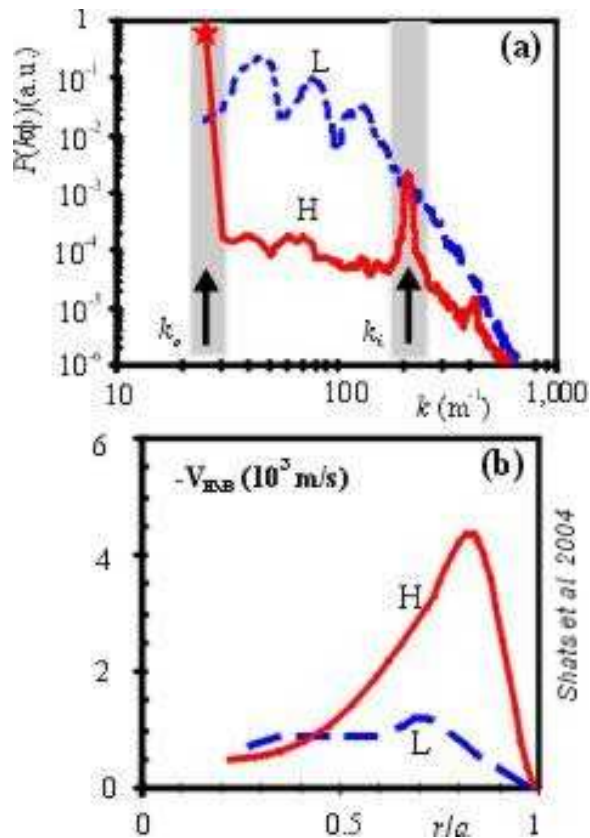}
\caption{\label{fig_3} Evolution of plasma turbulence during the
L-H transition. (a) Power spectrum of the electrostatic potential fluctuations in L (dashed line)
and H (solid) modes. The condensates maximum, denoted as a star, is
estimated from the radial velocity profile. (b) Radial profiles of
the $\textbf{E}\times\textbf{B}$ drift velocity,
$V_{\textbf{E}\times\textbf{B}}$, in two confinement modes.}
\end{figure}

We study the evolution of turbulence spectra in H-1 in the range of plasma parameters where transitions from low to high (L-H) confinement modes are observed \cite{Sha96}. The spectra are computed from the time series of the plasma electrostatic potential and then are rescaled into the wave number domain as $k=(2\pi f / V_{E \times B})$, where $f$ is the frequency and $V_{E \times B}$ is the ${E \times B}$ drift velocity. This is justified due to the linear $k(f)$ dependence \cite{Xia03}. In low confinement (L) mode the spectrum of potential fluctuations shows low-frequency coherent structures in the low wave-number range, and the decaying broadband turbulence in the higher-$k$ range (Fig.~\ref{fig_3}). The spectral energy transfer analysis \cite{Xia04} indicates that the energy source is localised in a spectral region at around $k_i \approx 200$~m$^{-1}$. This is the range of the underlying linear instability. The energy reservoir for this unstable range is the plasma pressure gradient. The broadband spectrum is generated through a \textit{random-phase} 3-wave interaction process, via the inverse cascade mechanism. The coherent structures are driven via a nonlocal coupling mechanism \cite{Xia04}, i.e. they obtain energy from the spectral region such that $k_c \ll k_1 \approx k_2 \approx k_i$, where $k_c$ is the wave number of the structure, $k_c=k_1-k_2$. In contrast to the broadband part of the spectrum, the nonlocal transfer occurs through \textit{coherent-phase} interactions.

The coherent structures have low poloidal mode numbers $m=1, 2$, etc. \cite{Sha99}. Recently we reported the first observation of the $m=0$ low-frequency (several kilohertz) mode, referred to as zonal flow \cite{Sha02}. These oscillating zonal flows have finite radial and nearly zero poloidal wave numbers, $k_r \gg k_{\theta} \approx 0$. They are driven by the unstable range via the nonlocal coherent spectral transfer \cite{Xia04}. A similar time-varying flow was observed in 2D fluid experiments. Sommeria \cite{Som86} reports a periodic reversal of the direction of the global flow, or the oscillating $m=0$ mode, in the regime characterised by a slightly higher damping than the one necessary for the establishment of a steady condensate regime. This suggests that oscillating zonal flows may be considered as an intermediate stage of the spectral condensation. The finite-frequency zonal flows play an important role in the plasma dynamics before the bifurcation to H-mode \cite{Pun04,Kim03}. The turbulent regime, dominated by these flows, has been likened to a metastable state in which the oscillating flows act as nuclei in L-H transitions (in analogy with water droplets in the supersaturated vapour before the phase transition) \cite{Pun04}.

During transitions from the L to H-mode, turbulent fluctuations are decreased in a broad spectral range. The only spectral region in which the fluctuation level is not reduced, is the one corresponding to the linear instability at about $k_i \approx 200$~m$^{-1}$ (Fig.~\ref{fig_3}(a)). The reduction in turbulent fluctuations coincides with the development of a strong mean \mbox{$E \times B$} flow in the plasma (Fig.~\ref{fig_3}(b)). The analogy with the 2D fluid experiment suggests that this flow develops as a result of the turbulence self-organization through the spectral condensation. Of course, part of the mean \mbox{$E \times B$} flow in H-mode may be generated by various transport mechanisms (neoclassical, particle orbit loss etc.). However the contribution of turbulence to the flow is dominant. We compute the turbulent energy contained in the spectrum of Fig.~\ref{fig_3}(a) before the transition,
\begin{equation}
W_T \sim \int_0^a rdr \Bigl( \frac{n_{eL}}{B^2} \sum_{k=0}^{k_i} k^2 \varphi_k^2 \Bigr)
\end{equation}
and compare it with the increase in the energy of the mean flow during the L-H transition,
\begin{equation}
\Delta W_{F} \sim \int_0^a rdr \Bigl( n_{eH}V_H^2 - n_{eL}V_L^2 \Bigr)
\end{equation}

\begin{figure}
\includegraphics[height=10.5 cm]{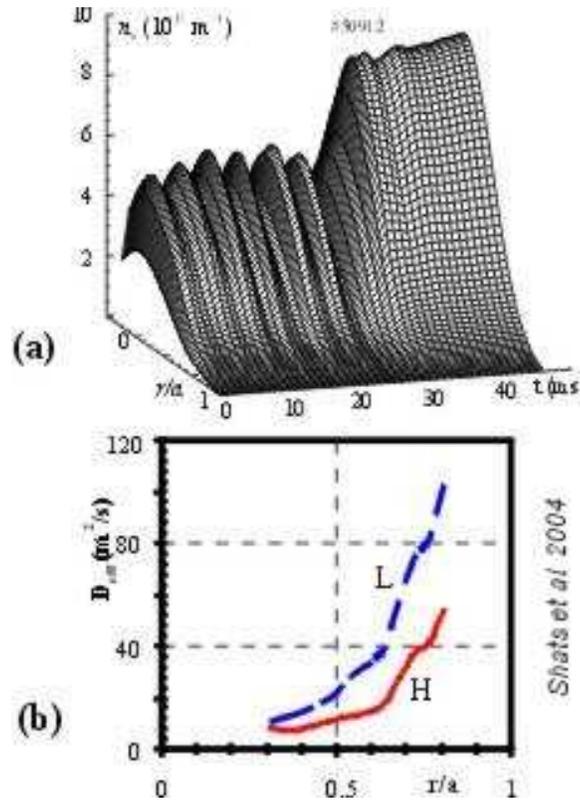}
\caption{\label{fig_4} Evolution of plasma parameters during the
confinement transition (multichannel spectroscopy diagnostic
data). (a) Time evolution of the electron density during
spontaneous L-H transition. (b) Radial profiles of the effective
diffusion coefficient in two confinement modes.}
\end{figure}

Here $n_{eL}$ and $n_{eH}$ are the electron densities in L and H modes respectively, and $B$ is the magnetic field. These energy estimates agree in our experiment within 20\%. This confirms that the observed increase in the \mbox{$E \times B$} flow in H-mode is largely due to the redistribution of the spectral energy from the intermediate scale into the mean zonal flow. After the flow establishes, it is sustained by a relatively small energy transfer from the unstable range, as discussed in \cite{Xia04}. The simultaneous reduction in turbulence and the generation of the flow during the transition from L to H mode coincide with a large increase in the electron density (and plasma pressure) and a substantial reduction of the particle diffusion in the plasma (Fig.~\ref{fig_4}).

The evolution of spectra in plasma across the L-H transition is remarkably similar to that in the 2D fluid experiment during spectral condensation (Fig.~\ref{fig_2}(b)). In both cases turbulent energy is reduced in the spectral range below the injection wave number, $k_i$, and it is increased at the largest scale allowed by the system, $k_c$. Also, both spectra in the enstrophy inertial range, at $k > k_i$, scale similar: $E \sim k^{-3.8}$ in the plasma and $E \sim k^{-3.6}$ in 2D fluid, as predicted theoretically \cite{Hort94}. The onset of the strong \mbox{$E \times B$} flow in the plasma H-mode is seen as the condensation of the turbulent energy, similar to the results in fluid.

Several features of the experimental picture presented here are in agreement with the theoretical conclusions of Balk, Zakharov and Nazarenko \cite{Balk90}. Using the hypothesis of the nonlocal inverse cascade, they predicted a reduction of the intermediate scales in the turbulent spectrum during spectral condensation into the zonal flow. The splitting of the turbulent $k$-spectrum into two unconnected components: (a) an intensive zonal flow and (b) the high-$k$ jet at the injection scale was also shown in \cite{Balk90}. Our experimental results confirm that the nonlocal spectral coupling of the unstable spectral range to the coherent condensate is responsible for the generation of the oscillating zonal flow in L mode. After the transition to H-mode and the stabilization of the zonal flow, the energy transferred from the unstable range is much smaller than the energy of the zonal flow, but is sufficient to overcome the dissipation at the condensate scale \cite{Balk90}, in agreement with our estimations of the spectral energy transfer in H-mode \cite{Xia04}. Our conclusion about the spectrally nonlocal nature of the process is based on the analysis of the energy transfer from the unstable range into coherent modes using the amplitude correlation technique. However the spectral energy transfer analysis via the 3-wave interactions does not clearly show that spectral energy is pumped into the coherent structures \cite{Xia04}. Possibly, two mechanisms are in work: 3-wave interactions generate the broadband turbulence, while 4-wave interactions are essential for the condensate generation \cite{Dyac96,McCar04}.

Summarising, we report the first experimental evidence of the similarity between self-organization in 2D fluid turbulence and in magnetically confined toroidal plasma. Transitions from turbulent state to the higher-order state occur similar in both systems and show: (1) the reduction of turbulence in a broad spectral range; (2) the generation of the strong stable flow as a result of the spectral energy redistribution; and (3) the reduction in diffusion in the condensate regime. Since the macroscopic features of the plasma state transitions described above are universal for improved confinement regimes in any magnetically confined plasma, it is natural to suggest that the mechanism of their generation is also universal.

\begin{acknowledgments}
\end{acknowledgments}

\end{document}